\documentclass{article}
\usepackage{geometry}    
\usepackage{amsfonts}    
\usepackage{amssymb}     
\usepackage[all]{xy}     
\usepackage{amsmath}     
\usepackage[dvips]{graphicx}       
\usepackage{color}
\usepackage{pst-tree}    
\usepackage{pstricks}

\newcommand{\go}[1]{\mathfrak {#1}}
\newcommand{\rrbr}{]\!]}
\newcommand{\llbr}{[\![}

\newtheorem{The}{Theorem}

\newtheorem{Lem}[The]{Lemma}
\newtheorem{Pro}[The]{Proposition}

\newtheorem{Rk}[The]{Remark}

\begin{document}

\begin{center}
\Large{Hochschild Cohomology of Cubic Surfaces}
\end{center}

\vspace{.4cm}
\begin{center}
Fr\'ed\'eric BUTIN\footnote{Universit\'e de Lyon, Universit\'e
Lyon 1, CNRS, UMR5208, Institut
Camille Jordan,
43 blvd du 11 novembre 1918, F-69622 Villeurbanne-Cedex, France,
email: butin@math.univ-lyon1.fr
}\vspace{.5cm}\\
\end{center}

\begin{small}
\noindent\textbf{\textsc{Abstract}}\\
\noindent We consider the polynomial algebra $\mathbb{C}[\mathbf{z}]:=\mathbb{C}[z_1,\,z_2,\,z_3]$ and the polynomial $f:=z_1^3+z_2^3+z_3^3+3qz_1z_2z_3$, where~$q\in \mathbb{C}$. Our aim is to compute the Hochschild homology and cohomology of the cubic surface $$\mathcal{X}_f:=\{\mathbf{z}\in\mathbb{C}^3\ /\ f(\mathbf{z})=0\}.$$
For explicit computations, we shall make use of a method suggested by M. Kontsevich. Then, we shall develop it in order to determine the Hochschild homology and cohomology by means of multivariate division and Groebner bases. Some formal computations with Maple are also used.\\
\end{small}\\

\begin{small}

\end{small}

\section{\textsf{Introduction and main results}}

\subsection{\textsf{Deformation quantization}}

\noindent Let us consider a mechanical system $(M,\,\mathcal{F}(M))$, where
$M$ is a Poisson manifold and $\mathcal{F}(M)$ the algebra of
regular functions on $M$. In
classical mechanics, we study the (commutative) algebra
$\mathcal{F}(M)$, that is
to say the observables of the classical system. The evolution of an observable $f$ is determined by the Hamiltonian equation $\dot{f}=\{f,\,H\}$, where $H\in\mathcal{F}(M)$ is the Hamiltonian of the system. But quantum
mechanics, where the physical system is described by a (non
commutative) algebra of operators on a Hilbert space, gives more
precise results than its classical analogue. Therefore it is important to be able to quantize
it.\\
One available method is geometric quantization, which allows us to construct
in an explicit way a Hilbert space and an algebra $\mathcal{C}$ of operators on
this space (see the book \cite{GRS07}), and to associate to every classical observable $f\in\mathcal{F}(M)$ a quantum one $\widehat{f}\in\mathcal{C}$. This
very interesting method presents the drawback of being seldom
applicable.\\
That is why other methods, such as asymptotic
quantization and deformation quantization, have been introduced.
The latter, described in 1978 by F. Bayen, M. Flato, C. Fronsdal,
A. Lichnerowicz and D. Sternheimer in \cite{BFFLS78}, is a good
alternative: instead of constructing an algebra of operators on a
Hilbert space, we define a formal deformation of $\mathcal{F}(M)$.
This is given by the algebra of formal power series
$\mathcal{F}(M)[[\hbar]]$, endowed with some associative, but not
always commutative, star-product:
\begin{equation}\label{def}
  f\ast
g=\sum_{j=0}^\infty m_j(f,\,g)\hbar^j
\end{equation}
where the maps $m_j$ are bilinear and where $m_0(f,\,g)=fg$. Then
quantization is given by the map $f\mapsto \widehat{f}$, where the
operator $\widehat{f}$ satisfies $\widehat{f}(g)=f\ast g$.\\

\noindent We can wonder in which cases a Poisson manifold admits such a quantization.
The answer was given by Kontsevich in \cite{K97} who
constructed a star-product on every Poisson manifold. Besides, he
proved that if $M$ is a smooth manifold, then the equivalence
classes of formal deformations of the zero Poisson bracket are in
bijection with equivalence classes of star-products. Moreover, as
a consequence of the Hochschild-Kostant-Rosenberg
theorem, every abelian star-product is equivalent to a trivial one.\\
In the case where $M$ is a singular algebraic variety, say
\[
M=\{\mathbf{z}\in \mathbb{C}^n\ /\ f(\mathbf{z})=0\},
\]
where $f$ belongs to $\mathbb{C}[\mathbf{z}]$,
we shall consider
the algebra of functions on $M$, i. e. the quotient algebra
$\mathbb{C}[\mathbf{z}]\,/\,\langle f\rangle$. So the above
mentioned result is not always valid. However, the deformations of
the algebra $\mathcal{F}(M)$, defined by the formula~(\ref{def}),
are always classified by the Hochschild cohomology of
$\mathcal{F}(M)$, and we are led to the study of the Hochschild
cohomology of $\mathbb{C}[\mathbf{z}]\,/\,\langle
f\rangle$.\\

\subsection{\textsf{Some recent related works}}

\noindent Several recent articles were devoted to the study of particular
cases, for Hochschild as well as for Poisson homology and cohomology:\\

\noindent C. Roger and P. Vanhaecke, in \cite{RV02},
 calculate the Poisson cohomology of the affine plane $\mathbb{C}^2$, endowed
with the Poisson bracket
$f\,\partial_1\wedge\partial_2$, where $f$ is a
homogeneous polynomial. They express it in terms of the number of
irreducible components of the singular locus $\{\mathbf{z}\in
\mathbb{C}^2\ /\ f(\mathbf{z})=0\}$.\\
M. Van den Bergh and A. Pichereau, in \cite{VB94},
\cite{P05} and \cite{P06}, are interested in the case where $f$ is a weighted homogeneous polynomial
with an isolated singularity at the origin. They compute the
Poisson homology and cohomology, which may be
expressed in particular in terms of the Milnor number of the space
$\mathbb{C}[z_1,\,z_2,\,z_3]\ /\ \langle
\partial_1f,\,\partial_2f,\,\partial_3f\rangle$.\\

\noindent In \cite{AFLS00}, Jacques Alev, Marco A. Farinati,
Thierry Lambre and Andrea L. Solotar establish a fundamental
result: they compute all the Hochschild homology and cohomology
spaces of $A_n(\mathbb{C})^G$, where $A_n(\mathbb{C})$ is the Weyl
algebra, for every finite subgroup $G$ of
$\mathbf{Sp}_{2n}\mathbb{C}$. It is an interesting and classical
question to compare the Hochschild homology and cohomology of
$A_n(\mathbb{C})^G$ with the Poisson homology and cohomology of
the ring of invariants $\mathbb{C}[\mathbf{x},\,\mathbf{y}]^G$,
which is a quotient algebra of the
form~$\mathbb{C}[\mathbf{z}]\,/\,\langle f_1,\dots,\,f_m\rangle$.\\
Finally, C. Fronsdal studies in \cite{FK07} Hochschild homology
of singular curves of the plane. Besides, the appendix
of this article gives another way to calculate the Hochschild
cohomology in the more general case of complete intersections.\\

\subsection{\textsf{Main results of the paper}}

\noindent Let us consider the polynomial algebra $\mathbb{C}[\mathbf{z}]:=\mathbb{C}[z_1,\,z_2,\,z_3]$ and the polynomial $$f:=z_1^3+z_2^3+z_3^3+3qz_1z_2z_3\in \mathbb{C}[\mathbf{z}],$$ where $q\in \mathbb{C}$. The surface defined by $f$, i.e. $\mathcal{X}_f:=\{\mathbf{z}\in\mathbb{C}^3\ /\ f(\mathbf{z})=0\}$, is called a cubic surface. The aim of this article is to compute the Hochschild homology and cohomology of this algebraic variety. Besides, we will consider the desingularisation of $\mathcal{X}_f$, and also compute its Hochschild homology and cohomology.\\
Let us denote by $A$ the quotient algebra $A:=\mathbb{C}[\mathbf{z}]\,/\,\langle f\rangle$.\\

\noindent The main result of the article is given by two theorems:\\

\begin{The}$\\$
Let $HH^p$ be the space of Hochschild cohomology of degree $p$. Then we have
$$HH^0=A,\ HH^1=\{\nabla f\wedge\mathbf{g}\,/\,\mathbf{g}\in A^3\}\oplus \mathbb{C}^8,\ HH^2=\{g\nabla f\,/\,g\in A\}\oplus \mathbb{C}^8,\ \textrm{and}\ \forall\ p\geq 3,\ HH^p=\mathbb{C}^{8}.$$
\end{The}

\begin{The}$\\$
Let $HH_p$ be the space of Hochschild homology of degree $p$. Then we have
$$HH_0=A,\ HH_1=\nabla f\wedge A^3,\ HH_2=A^3\,/\,(\nabla f\wedge A^3),\ \textrm{and}\ \forall\ p\geq 3,\ HH_p=\mathbb{C}^{8}.$$
\end{The}

\noindent For explicit computations, we shall make use of a method suggested by M. Kontsevich in the appendix
 of \cite{FK07}. Then, we shall develop it in order to determine the Hochschild homology and cohomology by means of multivariate division and Groebner bases. Some formal computations with the software Maple will be used all along our study.\\

\subsection{\textsf{Outline of the paper}}

\noindent In Section 2 we state the main theorems about Groebner bases and regular sequences which are useful for our computations. Then we recall the classical definitions about Hochschild homology and cohomology, and the important results about deformations of associative algebras.\\
Section 3 is the main section of the article: we compute the Hochschild cohomology of $\mathcal{X}_f$. We begin by giving some results about the Koszul complex, then we describe the cohomology spaces, and we compute them by means of Groebner bases and formal computations.\\
Finally, in Section 4, we study the case of Hochschild homology of cubic surfaces.\\

\section{\textsf{Framework and tools}}

\subsection{\textsf{Groebner bases and regular sequences}}

\noindent For each ideal $J$ of
$\mathbb{C}[\mathbf{z}]$, we denote by $J_A$ the image of $J$ by
the canonical projection
\[\mathbb{C}[\mathbf{z}]\rightarrow A=\mathbb{C}[\mathbf{z}]/\langle
f_1,\dots,\ f_m\rangle.\] Similarly if $(g_1,\dots,\ g_r)\in A^r$
we denote by $\langle g_1,\dots,\ g_r\rangle _A$ the ideal of $A$
generated by
$(g_1,\dots,\ g_r)$.\\
Besides, if $g\in \mathbb{C}[\mathbf{z}]$, and if $J$ is an ideal
of $\mathbb{C}[\mathbf{z}]$, we set \[Ann_J(g):=J : \langle g\rangle=\{h\in
\mathbb{C}[\mathbf{z}]\ /\ hg=0\mod J\}.\] In particular, $g$
does not divide $0$ in
$\mathbb{C}[\mathbf{z}]/J$ if and only if $Ann_J(g)=J$.\\
Finally, we denote by $\nabla g$ the gradient of a polynomial
$g\in \mathbb{C}[\mathbf{z}]$, and by $\langle\nabla g\rangle$ the ideal generated by the three partial derivatives of $g$.\\
Moreover, we use the notation $\partial_j$ for the
partial derivative with respect to $z_j$.\\

\noindent Here, we recall some important results about Groebner bases (see \cite{RSP02} for more details). For $g\in\mathbb{C}[\mathbf{z}]$, we denote by $lt(g)$ its leading term (for the lexicographic order~$\preceq$). Given a sequence of polynomials $[g_1,\dots,\,g_k]$, we say that $g$ reduced with respect to $[g_1,\dots,\,g_k]$, if $q$ is zero or if no one of the terms of $q$ is divisible by one of the elements $lt(g_1),\dots,\,lt(g_r)$.\\
Given a non trivial ideal~$J$ of~$\mathbb{C}[\mathbf{z}]$, a \emph{Groebner basis} of $J$ is a finite subset $G$ of $J\backslash\{0\}$ such that for every~~\mbox{$f\in
J\backslash\{0\}$,} there exists $g\in G$ such that $lt(g)$ divides $lt(f)$.\\

\begin{Pro}\label{propGr}$\\$
Let $G$ be a Groebner basis of an ideal $J$ of $\mathbb{C}[\mathbf{z}]$. Then every element of $J$ that is reduced with respect to $G$, is zero.\\
For every $f\in\mathbb{C}[\mathbf{z}]$, there exists a unique polynomial $r_G(f)\in\mathbb{C}[\mathbf{z}]$, reduced with respect to $G$, such that $f\equiv r_G(f)\mod J$.
\end{Pro}

\noindent The unique polynomial $r_G(f)$ defined above is called the normal form of $f$. So we have $$\forall\ f\in \mathbb{C}[\mathbf{z}],\ f\in J\Leftrightarrow r_G(f)=0.$$

\noindent The following theorem will be usefull fo the computation of the Hochschild cohomology of cubic surfaces. Given $J$ a non trivial ideal of $\mathbb{C}[\mathbf{z}]$,
the \emph{set of $J-$standard terms}, is the set of monomials of $\mathbb{C}[\mathbf{z}]$ except the set of dominant terms of non zero elements of $J$.\\

\begin{The}\label{Macaulay}(Macaulay)$\\$
The set of $J-$standard terms is a basis of the quotient $\mathbb{C}-$vector space $\mathbb{C}[\mathbf{z}]\,/\,J$.
\end{The}

\noindent Let $A=\bigoplus_{d=0}^\infty A_d$ a graded $\mathbb{C}-$algebra of Krull dimension $n$ (i.e. the maximal number of elements of $A$ that are algebraically independent on $\mathbb{C}$ is $n$).\\
Let $H(A_+)$ be the set of elements of $A$ that are homogeneous of positive degree. A sequence $[\theta_1,\dots,\,\theta_n]$ of elements of $H(A_+)$ is a \emph{homogeneous system of parameters} if $A$ is a module of finite type on the ring $\mathbb{C}[\theta_1,\dots,\,\theta_n]$. In particular, the elements $\theta_1,\dots,\,\theta_n$ are algebraically independent.\\
According to the normalization lemma of Noether\footnote{If $A$ is a $\mathbb{C}-$algebra of finite type, then there exists some elements $x_1,\dots,\,x_n\in A$, algebraically independent on $\mathbb{C}$, such that $A$ is integer on $\mathbb{C}[x_1,\dots,\,x_n]$.}, a homogeneous system of parameters always exists.\\

\noindent Let us now introduce a technical tool: a sequence $[\theta_1,\dots,\,\theta_n]$ of $n$ elements of $A$ is called a \emph{regular sequence} if, for every $j\in\llbr1,\,n\rrbr$, $\theta_j$ is not a divisor of zero of $A\,/\,\langle \theta_1,\dots,\,\theta_{j-1}\rangle$.\\
If the elements $\theta_1,\dots,\,\theta_n$ are algebraically independent (and this is the case in particular if $[\theta_1,\dots,\,\theta_n]$ is a homogeneous system of parameters), then $[\theta_1,\dots,\,\theta_n]$ is a regular sequence if and only if $A$ is a free module on the algebra~$\mathbb{C}[\theta_1,\dots,\,\theta_n]$.\\

\begin{The}\label{Mac}(see \cite{St93})$\\$
Let $A$ be a graded $\mathbb{C}-$algebra, and $[\theta_1,\dots,\,\theta_n]$ a homogeneous system of parameters. The two following points are equivalent:\\
$\bullet$ $A$ is a \emph{free} module of finite type on $\mathbb{C}[\theta_1,\dots,\,\theta_n]$, i.e.
\begin{equation}\label{Hironaka}
 A=\bigoplus_{i=1}^m \eta_i \mathbb{C}[\theta_1,\dots,\,\theta_n],
\end{equation}
$\bullet$ For every homogeneous system of parameters $[\phi_1,\dots,\,\phi_n]$, $A$ is a \emph{free} module of finite type on~$\mathbb{C}[\phi_1,\dots,\,\phi_n]$.\\
\end{The}

\noindent The meaning of this theorem is the following: in a graded $\mathbb{C}-$algebra, if there exists a homogeneous system of parameters that is a regular sequence, then every homogeneous system of parameters is a regular sequence. In particular, if $f\in\mathbb{C}[z_1,\,z_2,\,z_3]$ is a weight homogeneous polynomial with an isolated singularity at the origin, then $[f,\,\partial_3f,\,\partial_2f]$ is a homogeneous system of parameters. Now $[z_1,\,z_2,\,z_3]$ is a homogeneous system of parameters that is trivially a regular sequence, therefore $[f,\,\partial_3f,\,\partial_2f]$ is a regular sequence.\\

\subsection{\textsf{Hochschild homology and cohomology and deformations of algebras}}

\noindent $\bullet$ Consider an associative $\mathbb{C}-$algebra,
denoted by $A$. The Hochschild cohomological complex of $A$ is
\[\xymatrix{\textsf{C}^0(A) \ar@{->}[r]^{d^{(0)}} &
\textsf{C}^1(A) \ar@{->}[r]^{d^{(1)}}
 & \textsf{C}^2(A) \ar@{->}[r]^{d^{(2)}}
 & \textsf{C}^3(A) \ar@{->}[r]^{d^{(3)}}
 & \textsf{C}^4(A) \ar@{->}[r]^{d^{(4)}}
 & \dots ,}\]
where the space $\textsf{C}^p(A)$ of $p-$cochains is defined by $\textsf{C}^p(A)=0$
for $p\in -\mathbb{N}^*$, $\textsf{C}^0(A)=A$, and for $p\in \mathbb{N}^*$,
$\textsf{C}^p(A)$ is the space of $\mathbb{C}-$linear maps from $A^{\otimes
p}$ to $A$. The differential $d=\bigoplus_{i=0}^\infty d^{(p)}$ is
given by
\[\begin{array}{rl}
  \forall\ C\in \textsf{C}^p(A), & \
d^{(p)}\,C(a_0,\dots,\,a_p) = a_0C(a_1,\dots,\,a_p) \\
   & \displaystyle{\hspace{1.5cm}-\sum_{i=0}^{p-1}(-1)^iC(a_0,\dots,\,a_ia_{i+1},\dots,\,a_p)
+(-1)^{p-1}C(a_0,\dots,\,a_{p-1})a_p.} \\
\end{array}\] We may write it in terms of
the Gerstenhaber bracket\footnote{Recall that for $F\in \textsf{C}^p(A)$
and $H\in \textsf{C}^q(A)$, the Gerstenhaber product is the element
$F\bullet H\in \textsf{C}^{p+q-1}(A)$ defined by $F\bullet
H(a_1,\dots,\,a_{p+q-1})=\sum_{i=0}^{p-1}(-1)^{i(q+1)}F(a_1,\dots,\,a_i,\,H(a_{i+1},\dots,\,a_{i+q}),\,a_{i+q+1},\dots,\,a_{p+q-1})$,
and the Gerstenhaber bracket is $[F,\,H]_G:=F\bullet
H-(-1)^{(p-1)(q-1)}H\bullet F$. See for example \cite{G63}, and
\cite{BCKT05} p. 38.} $[\cdot,\cdot]_G$ and of the product $\mu$
of $A$, as follows: \[d^{(p)}C=(-1)^{p+1}[\mu,\,C]_G.\] Then we
define the Hochschild cohomology of $A$ as the cohomology of the
Hochschild cohomological complex associated to $A$, i. e.
$HH^0(A):=\textrm{Ker}\ d^{(0)}$ and
 for $p\in \mathbb{N}^*,\ HH^p(A):=\textrm{Ker}\ d^{(p)}\ /\ \textrm{Im}\ d^{{(p-1)}}$.\\

\noindent $\bullet$ Similarly, the Hochschild homological complex of $A$ is
\[\xymatrix{\dots \ar@{->}[r]^{d_5} & \textsf{C}_4(A)\ar@{->}[r]^{d_4} &
\textsf{C}_3(A) \ar@{->}[r]^{d_3}
 & \textsf{C}_2(A) \ar@{->}[r]^{d_2}
 & \textsf{C}_1(A) \ar@{->}[r]^{d_1}
 & \textsf{C}_0(A),}\]
where the space of $p-$chains is given by $\textsf{C}_p(A)=0$ for $p\in
-\mathbb{N}^*$, $\textsf{C}_0(A)=A$, and for~$p\in \mathbb{N}^*$,
$\textsf{C}_p(A)=A\otimes A^{\otimes p}$. The differential
$d=\bigoplus_{i=0}^\infty d_p$ is given by
\[\begin{array}{l}
  d_p\,(a_0\otimes a_1\otimes\dots\otimes a_p) =a_0a_1\otimes
a_2\otimes\dots\otimes a_p \\
    \displaystyle{\hspace{2cm}+\sum_{i=1}^{p-1}(-1)^i a_0\otimes
a_1\otimes\dots \otimes a_i a_{i+1}\otimes\dots \otimes a_p
+(-1)^p a_p a_0\otimes a_1\otimes\dots \otimes a_{p-1}.} \\
\end{array}\] We
define the Hochschild homology of $A$ as the homology of the
Hochschild homological complex associated to $A$, i. e.
$HH_0(A):=A\,/\,\textrm{Im}\ d_1$ and
 for $p\in \mathbb{N}^*,\ HH_p(A):=\textrm{Ker}\ d_p\ /\ \textrm{Im}\ d_{p+1}$.\\

\noindent $\bullet$ We denote by $\mathbb{C}[[\hbar]]$ (resp.
$A[[\hbar]]$) the algebra of formal power series in the parameter
$\hbar$, with coefficients in $\mathbb{C}$ (resp. $A$). A
deformation, or star-product, of the algebra $A$ is a map $\ast$ from $A[[\hbar]]\times
A[[\hbar]]$ to $A[[\hbar]]$ which is
$\mathbb{C}[[\hbar]]-$bilinear and such that
\[\begin{array}{l} \forall\ k\in\mathbb{C},\ \forall\ s\in A[[\hbar]],\ \ k\ast s=s\ast k=s,\\
\forall\ (s,\,t)\in
A[[\hbar]]^2,\ \ s\ast t= st \mod \hbar A[[\hbar]], \\
\forall\ (s,\,t,\,u)\in
A[[\hbar]]^3,\ \ s\ast (t\ast u)=(s\ast t)\ast u.\\
\end{array}\]\\
This means that there exists a sequence of bilinear\footnote{In the definition of a star-product, we often assume that the bilinear maps $C_j$ are
bidifferential operators (see for example \cite{BCKT05}). Here, we do not make this assumption.} maps $C_j$
from $A\times A$ to $A$ of which the first term $C_0$ is the
multiplication of $A$ and such that \[\begin{array}{l}
  \displaystyle{\forall\ (a,\,b)\in A^2,\
a\ast b=\sum_{j=0}^{\infty}C_j(a,\,b)\hbar^j}, \\
  \displaystyle{\forall\ n\in \mathbb{N},\ \sum_{i+j=n}C_i(a,\,C_j(b,\,c))=\sum_{i+j=n}C_i(C_j(a,\,b),\,c),\ \textrm{i.e.}\
  \sum_{i+j=n}[C_i,\,C_j]_G=0.}\\
\end{array}\]
We say that the deformation is of order $p$ if the previous
formulae are satisfied (only) for $n\leq p$.\\

\noindent $\bullet$ The Hochschild cohomology plays an important
role in the study of deformations of the algebra $A$, by helping
us to classify them. In fact, if $\pi\in \textsf{C}^2(A)$, we may construct
a first order deformation $m$ of $A$ such that $m_1=\pi$ if and
only if $\pi\in \textrm{Ker}\,d^{(2)}$. Moreover, two first order
deformations are equivalent\footnote{Two deformations
$\displaystyle{m=\sum_{j=0}^{p}C_j\,\hbar^j}$ (with $C_j\in \textsf{C}^2(A)$) and
$\displaystyle{m'=\sum_{j=0}^{p}C'_j\,\hbar^j}$ (with $C'_j\in \textsf{C}^2(A)$) of order $p$
are called equivalent if there exists a sequence of linear maps
$\varphi_j$ from $A$ to $A$ of which the first term $\varphi_0$ is
the identity of $A$ and such that \[\begin{array}{l}
  \displaystyle{\forall\ a\in A,\
\varphi(a)=\sum_{j=0}^{\infty}\varphi_j(a)\hbar^j}, \\
  \displaystyle{\forall\ n\in \mathbb{N},\ \sum_{i+j=n}\varphi_i(C_j(a,\,b))=\sum_{i+j+k=n}C'_i(\varphi_j(a),\,\varphi_k(b)).}\\
\end{array}\]} if and only if their
difference is an element of $\textrm{Im}\,d^{(1)}$. So the set of
equivalence classes of first order deformations is
in bijection with $HH^2(A)$.\\
If $\displaystyle{m=\sum_{j=0}^{p}C_j\,\hbar^j}$ (with $C_j\in \textsf{C}^2(A)$)
is a deformation of order $p$, then we may extend $m$ to a
deformation of order $p+1$ if and only if there exists $C_{p+1}\in \textsf{C}^2(A)$
such that \[\begin{array}{c}
\displaystyle{\forall\ (a,\,b,\,c)\in A^3,\ \sum_{i=1}^p\left(C_i(a,\,C_{p+1-i}(b,\,c))-C_i(C_{p+1-i}(a,\,b),\,c)\right)=-d^{(2)}\,C_{p+1}(a,\,b,\,c),} \\
\\
\textrm{i.e.}\ \displaystyle{\sum_{i=1}^p[C_i,\,C_{p+1-i}]_G=2\,d^{(2)}C_{p+1}.} \\
\end{array}\]
\noindent According to the graded Jacobi identity for
$[\cdot,\cdot]_G$, the last sum belongs to
$\textrm{Ker}\,d^{(3)}$. So~$HH^3(A)$ contains the obstructions
to extend a deformation of order $p$ to a deformation of order $p+1$.\\

\subsection{\textsf{The Hochschild-Kostant-Rosenberg theorem}}

\noindent Let us now recall a fundamental result about the Hochschild cohomology
of a smooth algebra. For more details, see the article \cite{HKR62} and the book \cite{Lo98}, Section 3.4.\\

\noindent Let $A$ be a $\mathbb{C}-$algebra. The
$A-$module $\Omega^1(A)$ of K\"{a}hler differentials is the $A$-module generated by the $\mathbb{C}-$linear symbols $da$ for $a\in A$, such that
$d(ab)=a(bd)+b(da)$.\\
For $n\geq 2$, the $A-$module of $n-$differential forms is the exterior product \mbox{$\Omega^n(A):=\Lambda^n\,\Omega^1(A)$.} By convention, we set $\Omega^0(A)=A$.\\
The antisymmetrization map $\varepsilon_n$ is defined by
$$\varepsilon_n(a_1\wedge\dots\wedge a_n)=\sum_{\sigma\in\go{S}_n}\varepsilon(\sigma)\,a_{\sigma^{-1}(1)}\otimes\dots\otimes a_{\sigma^{-1}(n)},$$
where $\varepsilon(\sigma)$ is the sign of the permutation
$\sigma$.\\
Let us denote by $HH_*(A)=\bigoplus_{n=0}^\infty HH_n(A)$ and $\Omega^*(A)$ the exterior algebra of $A$. \\

\begin{The}\label{HKR}(Hochschild-Kostant-Rosenberg)$\\$
Let $A$ be a smooth algebra. Then the antisymmetrization map
$$\varepsilon_*:\Omega^*(A)\rightarrow HH_*(A)$$ is an isomorphism of graded algebras.\\
\end{The}

\section{\textsf{Hochschild cohomology of $\mathcal{X}_f$}}

\subsection{\textsf{Koszul complex}}

\noindent In the following, we shall use the results about the Koszul
complex recalled below (see the appendix of \cite{FK07}).\\

\noindent $\bullet$ We consider $(f_1,\,\dots,\,f_m)\in \mathbb{C}[\mathbf{z}]^m$, and
we denote by $A$ the quotient $\mathbb{C}[\mathbf{z}]\,/\,\langle f_1,\ \dots,\ f_m\rangle$.\ We assume that we have a \emph{complete
intersection}, i.e. the dimension of the set of solutions of the
system $$\{\mathbf{z}\in \mathbb{C}^n\ /\ f_1(\mathbf{z})=\dots=f_m(\mathbf{z})=0\}$$ is $n-m$.\\

\noindent $\bullet$ We consider the differential graded algebra
\[\Phi=A[\varepsilon_1,\dots,\,\varepsilon_n;\,u_1,\dots,\,u_m]=\frac{\mathbb{C}[z_1,\ \dots,\ z_n]}{\langle f_1,\ \dots,\
f_m\rangle}[\varepsilon_1,\dots,\,\varepsilon_n;\,u_1,\dots,\,u_m],\] where
$\varepsilon_i:=\frac{\partial}{\partial z_i}$ is an odd variable (i. e.
the $\varepsilon_i$'s anticommute), and $u_j$ an even
variable (i. e. the $u_j$'s commute).\\
The algebra $\Phi$ is endowed with the differential
\[\displaystyle{\delta=\sum_{j=1}^n \sum_{i=1}^m
\frac{\partial f_i}{\partial z_j}u_i\frac{\partial}{\partial
\varepsilon_j}},\] and the Hodge grading, defined by $deg(z_i)=0,\
deg(\varepsilon_i)=1,\ deg(u_j)=2.$\\

\noindent The complex associated to $\Phi$ is as follows:

\[\xymatrix{\Phi(0) \ar@{->}[r]^{\widetilde{0}} &
\Phi(1) \ar@{->}[r]^{\delta^{(1)}}
 & \Phi(2) \ar@{->}[r]^{\delta^{(2)}}
 & \Phi(3) \ar@{->}[r]^{\delta^{(3)}}
 & \Phi(4) \ar@{->}[r]^{\delta^{(4)}}
 & \dots }\]\\

\noindent We may now state the main theorem which will allow us to
calculate the Hochschild cohomology:

\begin{The}\label{kont}(Kontsevich)$\\$
Under the previous assumptions, the Hochschild cohomology of $A$
is isomorphic to the cohomology of the complex $(\Phi,\
\delta)$ associated with the differential graded algebra~$\Phi$.\\
\end{The}

\begin{Rk}$\\$
Theorem \ref{kont} may be seen as a generalization of the
Hochschild-Kostant-Rosenberg theorem to the case of
non-smooth spaces.\\
\end{Rk}

\noindent Set $H^0:=A$,\ $H^1:=\textrm{Ker}\,\delta^{(1)}$ and for
$j\geq 2$, $H^p:=\textrm{Ker}\,\delta^{(p)}\,/\,\textrm{Im}\,\delta^{(p-1)}$.\\
According to Theorem \ref{kont}, we have, for $p\in \mathbb{N}$,
$HH^p(A)\simeq H^p$.\\

\subsection{\textsf{Cohomology spaces}}

\noindent $\bullet$ We consider now the case
$A:=\mathbb{C}[z_1,\,z_2,\,z_3],/\,\langle f\rangle$ and we want
to calculate its Hochschild cohomology. The
different spaces of the complex are now given by
$$\begin{array}{l}
  \Phi(0)=A,\ \ \Phi(1)=A\varepsilon_1\oplus A\varepsilon_2\oplus A\varepsilon_3 \\
  \forall\ p\in \mathbb{N}^*,\ \Phi(2p)=A
u_1^p\oplus A u_1^{p-1}\varepsilon_1\varepsilon_2\oplus A
u_1^{p-1}\varepsilon_2\varepsilon_3\oplus A u_1^{p-1}\varepsilon_3\varepsilon_1,\\
\forall\ p\in \mathbb{N}^*,\ \Phi(2p+1)=A u_1^p
\varepsilon_1\oplus A u_1^p\varepsilon_2\oplus A u_1^p\varepsilon_3\oplus A u_1^{p-1}\varepsilon_1\varepsilon_2\varepsilon_3.
\end{array}$$

\noindent This defines the bases $\mathcal{B}_{p}$. We have
$\frac{\partial}{\partial\varepsilon_1}(\varepsilon_1\wedge\varepsilon_2\wedge\varepsilon_3)=1\wedge\varepsilon_2\wedge\varepsilon_3=\varepsilon_2\wedge\varepsilon_3\wedge
1$, thus $\delta^{(3)}(\varepsilon_1\varepsilon_2\varepsilon_3)=
\frac{\partial f}{\partial z_1} u_1\varepsilon_2\varepsilon_3+\frac{\partial
f}{\partial z_2} u_1\varepsilon_3\varepsilon_1
+\frac{\partial f}{\partial z_3} u_1\varepsilon_1\varepsilon_2.$\\

\noindent We set $Df:=\left(%
\begin{array}{ccc}
  \partial_3f & \partial_1f & \partial_2f \\
\end{array}%
\right)$. The matrices $[\delta^{(j)}]$ of $\delta^{(j)}$ in the bases $\mathcal{B}_{j},\ \mathcal{B}_{j+1}$ are therefore given by
\[\begin{array}{c}
[\delta^{(1)}]=\left(
\begin{array}{c}
  ^t\nabla f \\
  \mathbf{0}_{3,3}\\
\end{array}
\right),\
\forall\ p\in \mathbb{N}^*,\
[\delta^{(2p)}]=\left(
\begin{array}{cccc}
0 & \partial_{2}f & 0 & -\partial_{3}f \\
  0 & -\partial_{1}f & \partial_{3}f & 0 \\
    0 & 0 & -\partial_{2}f & \partial_{1}f \\
      0 & 0 & 0  & 0\\
\end{array}
\right),\
[\delta^{(2p+1)}]=\left(
\begin{array}{cc}
^t\nabla f & 0 \\
  \mathbf{0}_{3,3}  & ^tDf\\
\end{array}
\right). \\
\end{array}\]

\noindent $\bullet$ We deduce\\

\noindent $\begin{array}{rcl}
  H^0 & = & A\,. \\
\end{array}$\\

\noindent $\begin{array}{rcl}
  H^1 & = & \{g_1\varepsilon_1+g_2\varepsilon_2+g_3\varepsilon_3\ /\ (g_1,\,g_2,\,g_3)\in A^3\
\textmd{and}\
g_1\,\partial_{1}f+g_2\,\partial_{2}f+g_3\,\partial_{3}f=0\}\\ \\
 & \simeq & \left\{\mathbf{g}=\left(
\begin{array}{c}
  g_1 \\
  g_2 \\
  g_3 \\
\end{array}
\right)\in A^3\ /\ \mathbf{g}\cdot\nabla f=0\right\}. \\
\end{array}$\vspace{.5cm}\\

\noindent $\begin{array}{rcl} H^{2} & = & \frac{\{g_0
u_1+g_3\varepsilon_1\varepsilon_2+g_1\varepsilon_2\varepsilon_3+g_2\varepsilon_3\varepsilon_1\ /\
(g_0,\,g_1,\,g_2\,g_3)\in A^4\ \textmd{and}\
g_3\,\partial_{2}f-g_2\,\partial_{3}f=g_1\,\partial_{3}f-g_3\,\partial_{1}f=g_2\,\partial_{1}f-g_1\,\partial_{2}f=0\}}
{\{(g_1\,\partial_{1}f+g_2\,\partial_{2}f+g_3\,\partial_{3}f)u_1\
/\ (g_1,\,g_2,\,g_3)\in A^3\}} \\ \\
    & \simeq & \left\{\mathbf{g}=\left(
\begin{array}{c}
 g_0\\
  g_1 \\
  g_2 \\
 g_3 \\
\end{array}
\right)\in A^4\ \Big/\ \nabla f\wedge \left(
\begin{array}{c}
  g_1 \\
  g_2 \\
 g_3 \\
\end{array}
\right)=0\right\}  \ \Bigg/\   \left\{\left(
\begin{array}{c}
  \mathbf{g}\cdot\nabla f \\
  \mathbf{0}_{3,1} \\
\end{array}
\right)\ /\ \mathbf{g}\in A^3\right\} \\ \\
    & \simeq & \frac{A}{\langle \partial_{1}f,\ \partial_{2}f,\ \partial_{3}f\rangle_A}\oplus \{\mathbf{g}\in A^3\ /\ \nabla f\wedge \mathbf{g}=0\} .\\
\end{array}$\vspace{.5cm}\\

\noindent For $p\geq 2,$\\
$\begin{array}{rcl}
  H^{2p} & = & \frac{\left\{g_0 u_1^p+g_3u_1^{p-1}\varepsilon_1\varepsilon_2+g_1u_1^{p-1}\varepsilon_2\varepsilon_3+g_2u_1^{p-1}\varepsilon_3\varepsilon_1\
\Big/\ \substack{(g_0,\,g_1,\,g_2\,g_3)\in A^4\ \textmd{and}\ \\
g_3\,\partial_{2}f-g_2\,\partial_{3}f=g_1\,\partial_{3}f-g_3\,\partial_{1}f=g_2\,\partial_{1}f-g_1\,\partial_{2}f=0}\right\}}
{\{(g_1\,\partial_{1}f+g_2\,\partial_{2}f+g_3\,\partial_{3}f)u_1^p+g_0(\partial_{3}f\,u_1^{p-1}\varepsilon_1\varepsilon_2
+\partial_{1}f\,u_1^{p-1}\varepsilon_2\varepsilon_3+\partial_{2}f\,u_1^{p-1}\varepsilon_3\varepsilon_1)\
/\ (g_0,\,g_1,\,g_2,\,g_3)\in A^3\}} \\ \\
    & \simeq & \begin{scriptsize}\left\{\mathbf{g}=\left(
\begin{array}{c}
 g_0\\
  g_1 \\
  g_2 \\
 g_3 \\
\end{array}
\right)\in A^4\ \Big/\ \nabla f\wedge \left(
\begin{array}{c}
  g_1 \\
  g_2 \\
 g_3 \\
\end{array}
\right)=0\right\}  \ \Bigg/\   \left\{\left(
\begin{array}{c}
  \mathbf{g}\cdot\nabla f \\
  g_0\,\partial_{1}f \\
  g_0\,\partial_{2}f \\
  g_0\,\partial_{3}f \\
\end{array}
\right)\ /\ \mathbf{g}\in A^3\ \textmd{and}\ g_0\in A\right\}\end{scriptsize} \\
\\
    & \simeq & \frac{A}{\langle \partial_{1}f,\ \partial_{2}f,\ \partial_{3}f\rangle_A}\oplus
     \frac{\{\mathbf{g}\in A^3\ /\ \nabla f\wedge \mathbf{g}=0\}}{\left\{g\nabla f\ /\ g\in A\right\}}.\\
\end{array}$\vspace{.5cm}\\

\noindent For $p\in \mathbb{N}^*,$\\
$\begin{array}{rcl}
  H^{2p+1} & = & \frac{\left\{g_1 u_1^p\varepsilon_1+g_2 u_1^p\varepsilon_2+g_3 u_1^p\varepsilon_3+g_0u_1^{p-1}\varepsilon_1\varepsilon_2\varepsilon_3\
\Big/\ \substack{(g_0,\,g_1,\,g_2\,g_3)\in A^4\ \textmd{and}\
g_1\,\partial_{1}f+g_2\,\partial_{2}f+g_3\,\partial_{3}f=0
\\ g_0\,\partial_{3}f=g_0\,\partial_{1}f=g_0\,\partial_{2}f=0}\right\}}
{\{(g_3\,\partial_{2}f-g_2\,\partial_{3}f)u_1^p\varepsilon_1+(g_1\,\partial_{3}f-g_3\,\partial_{1}f)u_1^p\varepsilon_2
+(g_2\,\partial_{1}f-g_1\,\partial_{2}f)u_1^p\varepsilon_3\ /\
(g_1,\,g_2,\,g_3)\in A^3\}} \\ \\
    & \simeq & \begin{scriptsize}\left\{\left(
\begin{array}{c}
 g_1\\
  g_2 \\
  g_3 \\
 g_0 \\
\end{array}
\right)\in A^4\ \Bigg/\ \begin{array}{c}
  \nabla f\cdot \left(
\begin{array}{c}
  g_1 \\
  g_2 \\
 g_3 \\
\end{array}
\right)=0 \\
  g_0\,\partial_{3}f=g_0\,\partial_{1}f=g_0\,\partial_{2}f=0 \\
\end{array}
        \right\} \ \Bigg/\   \left\{\left(
\begin{array}{c}
  \nabla f \wedge \mathbf{g} \\
  0 \\
\end{array}
\right)\ \Big/\ \mathbf{g}\in A^3\right\} \end{scriptsize} \\ \\
    & \simeq & \frac{\left\{\mathbf{g}\in A^3\ / \ \nabla f\cdot \mathbf{g}=0\right\}}
    {\left\{\nabla f\wedge \mathbf{g}\ /\ \mathbf{g}\in A^3\right\}}\oplus\{g\in A\ /\
    g\,\partial_{3}f=g\,\partial_{1}f=g\,\partial_{2}f=0\}. \\
\end{array}$\vspace{.5cm}\\

\noindent The following section will allow us to make those various spaces more explicit.\\

\subsection{\textsf{Hochschild cohomology of $\mathcal{X}_f$}}

\noindent The three partial derivative of $f$ are $\partial_1f=3z_1^2+3qz_2z_3$,  $\partial_2f=3z_2^2+3qz_1z_3$, and $\partial_3f=3z_3^2+3qz_1z_2$. So, according to Euler's formula, we get $z_1\partial_1f+z_2\partial_2f+z_3\partial_3f=3f$, hence the inclusion $\langle f\rangle\subset\langle \partial_1f,\,\partial_2f,\,\partial_3f\rangle$. We immediately deduce the isomorphism:
$$\frac{A}{\langle\nabla f \rangle_A}\simeq \frac{\mathbb{C}[\mathbf{z}]}{\langle\nabla f \rangle}.$$
The following lemma gives an explicit expression of this quotient.\\

\begin{Lem}\label{lemme0}$\\$
The vector space $\mathbb{C}[\mathbf{z}]\,/\,\langle\nabla f \rangle$ is isomorphic to $$\textrm{Vect}(1,\,z_3,z_3^2,z_3^3,\,z_2,\,z_2z_3,\,z_2^2,\,z_1).$$
\end{Lem}

\underline{Proof:}\\
We use Maple: a Groebner basis of the ideal $\langle\nabla f \rangle$ is $$[z_3^4,\ z_3^2z_2,\ z_2^2z_3,\ -z_3^3+z_2^3,\ z_2^2+qz_1z_3,\ z_3^2+qz_1z_2,\ z_1^2+qz_2z_3].$$ So the lemma results from Macaulay's theorem. $\blacksquare$\\

\begin{Lem}\label{lemme1}$\\$
The set of solutions of the equation $p\partial_1f=0$ in the quotient algebra $\mathbb{C}[\mathbf{z}]\,/\,\langle f,\,\partial_2f,\,\partial_3f \rangle$ is the vector space $$\textrm{Vect}(z_1,\,z_1^2,z_2^2,z_2^2z_3,\,z_3^2,\,z_3^2z_2,\,z_2^2z_3^2,\,z_3^3).$$
\end{Lem}

\underline{Proof:}\\
We use Maple and the multivariate division. A Groebner basis of the ideal $\langle f,\,\partial_2f,\,\partial_3f \rangle$ is
$$[z_3^4,\ z_3^3z_2,\ -z_3^3+z_2^3,\ z_2^2+qz_1z_3,\ z_3^2+qz_1z_2,\ -z_3^3+z_1^3].$$
So, according to Macaulay's theorem, a basis of the quotient space is $$(1,\ z_1,\ z_1^2,\ z_2,\ z_2^2,\ z_3,\ z_2z_3,\ z_2^2z_3,\ z_3^2,\ z_3^2z_2,\ z_2^2z_3^2,\ z_3^3).$$
Let $p:=a_{{1}}+a_{{2}}z_{{1}}+a_{{3}}{z_{{1}}}^{2}+a_{{4}}z_{{2}}+a_{{5}}{z_{
{2}}}^{2}+a_{{6}}z_{{3}}+a_{{7}}z_{{2}}z_{{3}}+a_{{8}}{z_{{2}}}^{2}z_{
{3}}+a_{{9}}{z_{{3}}}^{2}+a_{{10}}{z_{{3}}}^{2}z_{{2}}+a_{{11}}{z_{{2}
}}^{2}{z_{{3}}}^{2}+a_{{12}}{z_{{3}}}^{3}$ an element of the quotient space. Thanks to the multivariate division by the Groebner basis, the normal form of $p\partial_1f$ with respect to the ideal $\langle f,\,\partial_2f,\,\partial_3f \rangle$ is $$3\,a_{{1}}{z_{{1}}}^{2}+3q\,a_{{7}}{z_{{2}}}^{2}{z_{{3}}}^{2}+3\,{
\frac {a_{{7}}}{{q}^{2}}}{z_{{2}}}^{2}{z_{{3}}}^{2}+3q\,a_{{4}}{z_{{2}}
}^{2}z_{{3}}+3\,{\frac {a_{{4}}}{{q}^{2}}}z_{{3}}{z_{{2}}}^{2}+3q\,a_{
{6}}{z_{{3}}}^{2}z_{{2}}+3\,{\frac {a_{{6}}}{{q}^
{2}}}z_{{2}}{z_{{3}}}^{2}+3q\,a_{{1}}z_{{2}}z_{{3}}.$$
So, $p\partial_1f$ belongs to the ideal $\langle f,\,\partial_2f,\,\partial_3f \rangle$ if and only if $a_1=a_4=a_6=a_7=0$. Hence the proof of the lemma. $\blacksquare$\\

\noindent We are now able to solve the equation $\mathbf{g}\cdot\nabla f=0$ in $A^3$. It is the object of the following lemma:

\begin{Lem}\label{lemme2}$\\$
The set of solutions of the equation $\mathbf{g}\cdot\nabla f=0$ in $A^3$ is isomorphic to the set $$\{\nabla f\wedge\mathbf{g}\,/\,\mathbf{g}\in A^3\}\oplus \mathbb{C}^8.$$
\end{Lem}

\underline{Proof:}\\
\noindent We have solve the equation $g\cdot\nabla f=0$ in $A$, i.e.
\begin{equation}\label{eq1}
    g_1\partial_1f+g_2\partial_2f+g_3\partial_3f=0\ \textrm{in}\ \mathbb{C}[\mathbf{z}]\,/\,\langle f\rangle.
\end{equation}
Let $g\in\mathbb{C}[\mathbf{z}]$ be a solution of $g\cdot\nabla f=0$, i.e. $g_1\partial_1f+g_2\partial_2f+g_3\partial_3f\in\langle f\rangle$.\\
Hence $g_1\partial_1f\in\langle f,\,\partial_2f,\,\partial_3f\rangle$. So, according to Lemma \ref{lemme1}, we have
\begin{equation}\label{eqetoile}
g_1=\varepsilon f+\eta\partial_2f+\zeta\partial_3f+\varphi,
\end{equation}
where $$\varphi=\sum_{i=1}^2a_iz_1^i+\sum_{i=0}^2 b_iz_2^2z_3^i+\sum_{i=0}^1 c_iz_3^2z_2^i+dz_3^3,$$ with $\varepsilon,\,\eta,\,\zeta\in \mathbb{C}[\mathbf{z}]$ and $a_i,\,b_i,\,c_i,\,d\in \mathbb{C}$.
Now, we have the Euler formula:
\begin{equation}\label{eq2}
z_1\partial_1f+z_2\partial_2f+z_3\partial_3f=3f.
\end{equation}
Moreover, we have the two following key-relations:
\begin{equation}\label{eq3}
 z_2^2\partial_1f=(z_1^2+2qz_2z_3)\partial_2f-3qz_3f+qz_3^2\partial_3f,
\end{equation}
\begin{equation}\label{eq4}
 z_3^2\partial_1f=(z_1^2+2qz_2z_3)\partial_3f-3qz_2f+qz_2^2\partial_2f,
\end{equation}
Thus $$\begin{array}{rcl}
        g_1\partial_1f & = & \displaystyle{\left(\sum_{i=1}^2a_iz_1^{i-1}\right)z_1\partial_1f+\left(\sum_{i=0}^2b_iz_3^{i}\right)z_2^2\partial_1f}\\
     & &  \displaystyle{+\left(dz_3+\sum_{i=0}^1c_iz_2^{i}\right)z_3^2\partial_1f+\varepsilon f\partial_1f+\eta\partial_1f\partial_2f+\zeta\partial_1f\partial_3f} \\
        & = & \displaystyle{\left(\sum_{i=1}^2a_iz_1^{i-1}\right)(3f-z_2\partial_2f-z_3\partial_3f)+\left(\sum_{i=0}^2b_iz_3^{i}\right)((z_1^2+2qz_2z_3)\partial_2f-3qz_3f+qz_3^2\partial_3f)}\\
      & & \displaystyle{+\left(dz_3+\sum_{i=0}^1c_iz_2^{i}\right)((z_1^2+2qz_2z_3)\partial_3f-3qz_2f+qz_2^2\partial_2f)+\varepsilon f\partial_1f+\eta\partial_1f\partial_2f+\zeta\partial_1f\partial_3f}.
      \end{array}
$$
Then, Equation (\ref{eq1}) becomes
$$\begin{array}{rcl}
 & &\displaystyle{\left(-\sum_{i=1}^2a_iz_1^{i-1}z_2+(z_1^2+2qz_2z_3)\sum_{i=0}^2b_iz_3^i+qz_2^2\left(dz_3+\sum_{i=0}^1c_iz_2^i\right)+g_2+\eta\partial_1f\right)\partial_2f}\\
 & &\displaystyle{\left(-\sum_{i=1}^2a_iz_1^{i-1}z_3+qz_3^2\sum_{i=0}^2b_iz_3^i+(z_1^2+2qz_2z_3)\left(dz_3+\sum_{i=0}^1c_iz_2^i\right)+g_3+\zeta\partial_1f\right)\partial_3f}\\
 & = & 0\ \textrm{in}\ \mathbb{C}[\mathbf{z}]\,/\,\langle f\rangle.
  \end{array}
$$
Hence $$\left(-\sum_{i=1}^2a_iz_1^{i-1}z_2+(z_1^2+2qz_2z_3)\sum_{i=0}^2b_iz_3^i+qz_2^2\left(dz_3+\sum_{i=0}^1c_iz_2^i\right)+g_2+\eta\partial_1f\right)\partial_2f=0\ \textrm{in}\ \mathbb{C}[\mathbf{z}]\,/\,\langle f,\,\partial_3f\rangle.$$
Now, as $\langle f,\,\partial_3f\rangle,\,\partial_2f$ is a regular sequence, we have $\textrm{Ann}_{\langle f,\,\partial_3f\rangle}(\partial_2f)=\langle f,\,\partial_3f\rangle\,:\,\langle\partial_2f\rangle=\langle f,\,\partial_3f\rangle,$
therefore $$-\sum_{i=1}^2a_iz_1^{i-1}z_2+(z_1^2+2qz_2z_3)\sum_{i=0}^2b_iz_3^i+qz_2^2\left(dz_3+\sum_{i=0}^1c_iz_2^i\right)+g_2+\eta\partial_1f\in
\mathbb{C}[\mathbf{z}]\,/\,\langle f,\,\partial_3f\rangle,$$
i.e.
\begin{equation}\label{eq2etoiles}
    g_2=\alpha f+\beta\partial_3f-\eta\partial_1f+\psi
\end{equation}
with $$\psi:=\sum_{i=1}^2a_iz_1^{i-1}z_2-(z_1^2+2qz_2z_3)\sum_{i=0}^2b_iz_3^i-qz_2^2\left(dz_3+\sum_{i=0}^1c_iz_2^i\right).$$
By substituting, we get 
\begin{equation}\label{eq5}
    \varphi\partial_1f+\psi\partial_2f+(\alpha\partial_2f+\varepsilon\partial_1f)f+(\beta\partial_2f+\zeta\partial_1f+g_3)\partial_3f=0\ \textrm{in}\ \mathbb{C}[\mathbf{z}]\,/\,\langle f\rangle.
\end{equation}
Now, according to (\ref{eq2}), (\ref{eq3}), (\ref{eq4}), we have $$\begin{array}{rcl}
         \varphi\partial_1f+\psi\partial_2f & = & \displaystyle{(z_2^2\partial_1f-(z_1^2+2qz_2z_3)\partial_2f)\sum_{i=0}^2b_iz_3^i+(z_3^2\partial_1f-qz_2^2\partial_2f)\left(dz_3+\sum_{i=0}^1c_iz_2^i\right)}\\
          & & \displaystyle{+(z_1\partial_1f+z_2\partial_2f)\sum_{i=1}^2a_iz_1^{i-1}} \\
          & = & \displaystyle{(qz_3^2\partial_3f-3qz_3f)\sum_{i=0}^2b_iz_3^i+((z_1^2+2qz_2z_3)\partial_3f-3qz_2f)\left(dz_3+\sum_{i=0}^1c_iz_2^i\right)}\\
         & &   \displaystyle{+(3f-z_3\partial_3f)\sum_{i=1}^2a_iz_1^{i-1}},
       \end{array}
$$
therefore
\begin{equation}\label{eq6}
\begin{array}{rcl}
  \varphi\partial_1f+\psi\partial_2f & = & \displaystyle{\left(qz_3^2\sum_{i=0}^2b_iz_3^i+(z_1^2+2qz_2z_3)\left(dz_3+\sum_{i=0}^1c_iz_2^i\right)-z\sum_3\sum_{i=1}^2a_iz_1^{i-1}\right)\partial_3f}\\
 & &     \displaystyle{+\left(-3qz_3\sum_{i=0}^2b_iz_3^i-3qz_2\left(dz_3+\sum_{i=0}^1c_iz_2^i\right)+3\sum_{i=1}^2a_iz_1^{i-1}\right)f}.
\end{array}
\end{equation}
By another substituting, we obtain 
$$\left(qz_3^2\sum_{i=0}^2b_iz_3^i+(z_1^2+2qz_2z_3)\left(dz_3+\sum_{i=0}^1c_iz_2^i\right)-z_3\sum_{i=1}^2a_iz_1^{i-1}+\zeta\partial_1f+\beta\partial_2f+g_3\right)\partial_3f=0\ \textrm{in}\ \mathbb{C}[\mathbf{z}]\,/\,\langle f\rangle.$$
As $\partial_3f$ and $f$ are coprime, we deduce
\begin{equation}\label{eq3etoiles}
    g_3=\gamma f-\zeta\partial_1f-\beta\partial_2f+\chi,
\end{equation}
with $\gamma\in \mathbb{C}[\mathbf{z}]$, where
$$\chi:=-qz_3^2\sum_{i=0}^2b_iz_3^i-(z_1^2+2qz_2z_3)\left(dz_3+\sum_{i=0}^1c_iz_2^i\right)+z_3\sum_{i=1}^2a_iz_1^{i-1}.$$
Now, let $\mathbf{g}$ be defined by the Equations (\ref{eqetoile}), (\ref{eq2etoiles}), (\ref{eq3etoiles}). Then
$$\mathbf{g}\cdot\nabla f=\varphi\partial_1f+\psi\partial_2f+\chi\partial_3f+(\varepsilon\partial_1f+\alpha\partial_2f+\gamma\partial_3f)f\in \langle f\rangle,$$
according to Relation (\ref{eq6}).\\
Finally, we have proved $$\{\mathbf{g}\in A^3\,/\,\mathbf{g}\cdot\nabla f=0\}=\left\{\nabla f\wedge\left(
                                                                                                     \begin{array}{c}
                                                                                                       \beta \\
                                                                                                       -\zeta \\
                                                                                                       \eta \\
                                                                                                     \end{array}
                                                                                                   \right)+\left(
                                                                                                             \begin{array}{c}
                                                                                                               \varphi \\
                                                                                                               \psi \\
                                                                                                               \chi \\
                                                                                                             \end{array}
                                                                                                           \right)
\right\},$$
hence $$\{\mathbf{g}\in A^3\,/\,\mathbf{g}\cdot\nabla f=0\}\simeq \{\nabla f\wedge\mathbf{g}\,/\,\mathbf{g}\in A^3\}\oplus \mathbb{C}^8.\ \blacksquare$$\\

\noindent So, the cohomology spaces of odd degrees are given by $$\begin{array}{rcl}
                                                         H^1 & \simeq & \{\nabla f\wedge\mathbf{g}\,/\,\mathbf{g}\in A^3\}\oplus \mathbb{C}^8,\\
                                                         \forall\ p\in \mathbb{N}^*,\  H^{2p+1} & \simeq & \mathbb{C}^8.
                                                       \end{array}
$$
We now determine the set  $\{\mathbf{g}\in A^3\,/\,\nabla f\wedge \mathbf{g}=0\}$:\\

\begin{Lem}\label{lemme3}$\\$
The set of solutions of the equation $\nabla f\wedge \mathbf{g}=0$ in $A$ is isomorphic to the set $\{g\nabla f\,/\,g\in A\}$.
\end{Lem}

\underline{Proof:}\\
Let $\mathbf{g}\in A^3$ be such that $\nabla f\wedge \mathbf{g}=0$ , i.e.
$$g_3\partial_2f-g_2\partial_3f=g_1\partial_3f-g_3\partial_1f=g_2\partial_1f-g_1\partial_2f=0 \mod \langle f\rangle.$$
Then
\begin{equation}\label{eq7}
 g_3\partial_2f-g_2\partial_3f=0 \mod \langle f\rangle,
\end{equation}
hence $g_3\partial_2f=0\mod \langle f,\,\partial_3f\rangle.$\\
As $\langle f,\,\partial_3f\rangle,\,\partial_2f$ is a regular sequence, we have $\langle f,\,\partial_3f\rangle : \langle \partial_2f\rangle=\langle f,\,\partial_3f\rangle$, thus $g_3\in\langle f,\,\partial_3f\rangle$, i.e. $g_3=\alpha f+\beta\partial_3f$.\\
We get $$(\alpha f+\beta\partial_3f)\partial_2f-g_2\partial_3f=0\ \textrm{in}\ \mathbb{C}[\mathbf{z}]\,/\,\langle f\rangle,$$  
hence $(\beta\partial_2f-g_2)\partial_3f=0\ \textrm{in}\ \mathbb{C}[\mathbf{z}]\,/\,\langle f\rangle$. As $f$ and $\partial_3f$ are coprime, we deduce $\beta\partial_2f-g_2\in\langle f\rangle$, i.e.~$g_2=\beta\partial_2f-\gamma f$.\\
By substituting in $g_1\partial_3f-g_3\partial_1f=0 \mod \langle f\rangle$, we obtain $(g_1-\beta \partial_1f)\partial_3f=0 \mod \langle f\rangle$.\\
Similarly, we have $g_1-\beta\partial_1f\in\langle f\rangle$, i.e.~$g_1=\beta\partial_1f+\varepsilon f$.\\ 
Finally, $$\{\mathbf{g}\in A^3\,/\,\nabla f\wedge \mathbf{g}=0\}=\{g\nabla f\,/\,g\in A\}.\ \blacksquare$$\\

\noindent So, the cohomology spaces of even degrees are
$$\begin{array}{rcl}
    H^2 & \simeq & \{g\nabla f\,/\,g\in A\}\oplus \mathbb{C}^8 \\
    \forall\ p\geq 2,\ H^{2p} & \simeq & \mathbb{C}^8
  \end{array}
$$

\noindent Finally, the Hochschild cohomology is given by the following theorem.

\begin{The}\label{Thecohom}$\\$
Let $HH^p$ be the space of Hochschild cohomology of degree $p$. Then we have
$$HH^0=A,\ HH^1=\{\nabla f\wedge\mathbf{g}\,/\,\mathbf{g}\in A^3\}\oplus \mathbb{C}^8,\ HH^2=\{g\nabla f\,/\,g\in A\}\oplus \mathbb{C}^8,\ \textrm{and}\ \forall\ p\geq 3,\ HH^p=\mathbb{C}^{8}.$$
\end{The}

\section{\textsf{Hochschild homology of $\mathcal{X}_f$}}

\subsection{\textsf{Koszul complex}}

\noindent $\bullet$ There is an analogous of Theorem \ref{kont}
for the Hochschild homology: we consider the complex
\[\Psi=A[\zeta_1,\dots,\,\zeta_n;\,v_1,\dots,\,v_m],\]
where $\zeta_i$ is an odd variable and $v_j$ an even variable.
$\Psi$ is endowed with the differential
\[\displaystyle{\vartheta=\sum_{i=1}^n
\sum_{j=1}^m \frac{\partial f_j}{\partial
z_i}\zeta_i\frac{\partial}{\partial v_j}},\] and the Hodge
grading,
defined by $deg(z_i)=0,\ deg(\zeta_i)=-1,\ deg(v_j)=-2.$\\

\[\xymatrix{\dots \ar@{->}[r]^{\vartheta^{(-5)}} &
\Psi(-4) \ar@{->}[r]^{\vartheta^{(-4)}}
 & \Psi(-3) \ar@{->}[r]^{\vartheta^{(-3)}}
 & \Psi(-2) \ar@{->}[r]^{\vartheta^{(-2)}}
 & \Psi(-1) \ar@{->}[r]^{\vartheta^{(-1)}}
 & \Psi(0) }\]

\begin{The}\label{konthom}(Kontsevich)$\\$
Under the previous assumptions, the Hochschild homology of $A$ is
isomorphic to the cohomology of the complex $(\Psi,\
\vartheta)$.
\end{The}

\noindent Set $L^0:=A\,/\,\textrm{Im}\,\vartheta^{(-1)}$,\ and for
$p\geq 1$, $L^{-p}:=\textrm{Ker}\,\vartheta^{(-p)}\,/\,\textrm{Im}\,\vartheta^{(-p-1)}$.\\
According to Theorem \ref{konthom}, we have, for $p\in
\mathbb{N}$,
$HH_p(A)\simeq L^{-p}$.\\

\subsection{\textsf{Hochschild homology of $\mathcal{X}_f$}}

\noindent $\bullet$ Here, we have \[\begin{array}{rl}
 &  \Psi(0)=A, \\
  &  \Psi(-1)=A\zeta_1\oplus A\zeta_2\oplus A\zeta_3, \\
  \forall\ p\in \mathbb{N}^*,  & \Psi(-2p)=Av_1^p\oplus Av_1^{p-1}\zeta_1\zeta_2\oplus Av_1^{p-1}\zeta_2\zeta_3\oplus Av_1^{p-1}\zeta_3\zeta_1, \\
  \forall\ p\in \mathbb{N}^*,  & \Psi(-2p-1)=Av_1^p\zeta_1\oplus Av_1^{p}\zeta_2\oplus Av_1^{p}\zeta_3\oplus Av_1^{p-1}\zeta_1\zeta_2\zeta_3. \\
\end{array}\]
This defines the bases $\mathcal{V}_{p}$. The differential is
$\vartheta=(\zeta_1\partial_1f+\zeta_2\partial_2f+\zeta_3\partial_3f)\frac{\partial}{\partial
v_1}$.\\

\noindent By setting $Df:=\left(%
\begin{array}{ccc}
  \partial_3f & \partial_1f & \partial_2f \\
\end{array}%
\right)$, we deduce the matrices $[\vartheta^{(-j)}]$ of $\vartheta^{(-j)}$ in the bases $\mathcal{V}_{-j},\ \mathcal{V}_{-j+1}$: \[\begin{array}{l}
  [\vartheta^{(-2)}]=\left(%
\begin{array}{cc}
  \nabla f & \mathbf{0}_{3,3} \\
\end{array}%
\right), \\
  \forall\ p\geq 2,\ [\vartheta^{(-2p)}]=\left(%
\begin{array}{cc}
  \nabla f & \mathbf{0}_{3,3} \\
  0 & (p-1)Df \\
\end{array}%
\right),\
  \forall\ p\geq 1,\ [\vartheta^{(-2p-1)}]=\left(%
\begin{array}{cccc}
  0 & 0 & 0 & 0 \\
  -p\partial_2f & p\partial_1f & 0 & 0 \\
  0 & -p\partial_3f & p\partial_2f & 0 \\
  p\partial_3f & 0 & -p\partial_1f & 0 \\
\end{array}%
\right). \\
\end{array}
\]

\noindent $\bullet$ The cohomology spaces read :\\

\noindent $\begin{array}{rcl}
  L^0 & = & A \\
\end{array},$\\

\noindent $\begin{array}{rcl}
  L^{-1} & = & \frac{A^3}{\{g\nabla f\ /\ g\in A\}} \\
\end{array},$\\

\noindent $\begin{array}{rcl}
  L^{-2} & = & \{g\in A\ /\ g\partial_1f=g\partial_2f=g\partial_3f=0\}\oplus \frac{A^3}{\{\nabla f\wedge \mathbf{g}\ /\ \mathbf{g}\in A^3\}} \\
\end{array}.$\\

\noindent For $p\geq 2$,\\
$\begin{array}{rcl}
  L^{-2p} & \simeq & \{g\in A\ /\ g\partial_1f=g\partial_2f=g\partial_3f=0\}\oplus \frac{\{\mathbf{g}\in A^3\ /\ \mathbf{g}\cdot\nabla f=0\}}{\{\nabla f\wedge \mathbf{g}\ /\ \mathbf{g}\in A^3\}} \\
\end{array}$.\\

\noindent For $p\in \mathbb{N}^*$,\\
$\begin{array}{rcl}
L^{-2p-1} & \simeq &  \frac{\{\mathbf{g}\in A^3\ /\ \nabla f\wedge\mathbf{g}=0\}}{\{g\nabla f\ /\ g\in A\}}\oplus \frac{A}{\langle \nabla f\rangle_A}\\
\end{array}$.\vspace{.4cm}\\

\noindent $\bullet$ We have
$\left\{g\in A\ /\
g\partial_1f=g\partial_2f=g\partial_3f=0\right\}=\{0\}$, and according to Euler's formula,~$\frac{A}{\langle\nabla
f\rangle_A}\simeq\frac{\mathbb{C}[\mathbf{z}]}{\langle\nabla
f\rangle}$.\\
With the exception of $A^3\,/\,\{\nabla f\wedge \mathbf{g}\,/\,\mathbf{g}\in A^3\}$, the spaces quoted above have been computed in the preceding section, in particular $A^3\,/\,A\nabla f\simeq \nabla f\wedge A^3$ (See Remark ).\\
Now $\{\nabla f\wedge \mathbf{g}\,/\,\mathbf{g}\in
A^3\}\subset\{\mathbf{g}\in A^3\,/\,\mathbf{g}\cdot\nabla f=0\}$,
therefore \[\dim\left(A^3\,/\,\{\nabla f\wedge
\mathbf{g}\,/\,\mathbf{g}\in A^3\}\right)\geq
\dim\left(A^3\,/\,\{\mathbf{g}\in A^3\,/\,\mathbf{g}\cdot\nabla
f=0\}\right),\] and $A^3\,/\,\{\mathbf{g}\in
A^3\,/\,\mathbf{g}\cdot\nabla f=0\}\simeq \{\mathbf{g}\cdot\nabla
f\,/\,\mathbf{g}\in A^3\}$. As the map
\[g\in A\mapsto \left(%
\begin{array}{c}
  g \\
  0 \\
  0 \\
\end{array}%
\right)\cdot\nabla f\in \{\mathbf{g}\cdot\nabla
f\,/\,\mathbf{g}\in A^3\}\] is injective, $A^3\,/\,\{\nabla
f\wedge \mathbf{g}\,/\,\mathbf{g}\in A^3\}$ is infinite dimensional.\vspace{.4cm}\\

\noindent $\bullet$ Finally, the Hochschild homology is given by the following theorem.

\begin{The}\label{Thehom}$\\$
Let $HH_p$ be the space of Hochschild homology of degree $p$. Then we have
$$HH_0=A,\ HH_1=\nabla f\wedge A^3,\ HH_2=A^3\,/\,(\nabla f\wedge A^3),\ \textrm{and}\ \forall\ p\geq 3,\ HH_p=\mathbb{C}^{8}.$$
\end{The}

\begin{small}

\end{small}


\begin{thebibliography}{99}
\footnotesize\itemsep=0pt

\bibitem{AFLS00}
Alev J., Farinati M.A., Lambre T., Solotar A. L., Homologie des
invariants d'une alg\`ebre de Weyl sous l'action d'un groupe fini,
{\it Journal of Algebra} {\bf 232} (2000), 564--577.

\bibitem{BCKT05}
Brugui\`eres A., Cattaneo A., Keller B., Torossian C.,
\textit{D\'eformation, Quantification, Th\'eorie de Lie},
 Panoramas et Synth\`eses, SMF, 2005.\\

\bibitem{BFFLS78}
Bayen F., Flato M., Fronsdal C., Lichnerowicz A., Sternheimer D.,
Deformation theory and quantization. I and II. Physical
applications, {\it Ann. Physics} {\bf111} (1978), no. 1, 61--110
and 111--151.


\bibitem{FK07}
Fronsdal C., Kontsevich M., Quantization on Curves, {\it Lett.
Math. Phys.} {\bf 79} (2007), 109--129, math-ph/0507021.

\bibitem{G63}
Gerstenhaber M., The cohomology structure of an associative ring,
{\it Annals of Math. (2)} \textbf{78} (1963), 267--288.

\bibitem{GRS07}
Guieu L., Roger C., avec un appendice de Sergiescu V., L'Alg\`ebre
et le Groupe de Virasoro: aspects g\'eom\'etriques et
alg\'ebriques, g\'en\'eralisations, Publication du Centre de
Recherches Math\'ematiques de Montr\'eal, s\'erie ``Monographies,
notes de cours et Actes de conf\'erences'', PM28, 2007.

\bibitem{Harr92}
Harris J., Algebraic Geometry, Springer-Verlag, 1992.

\bibitem{Hart77}
Hartshorne R., Algebraic Geometry, Springer-Verlag, 1977.

\bibitem{HKR62}
Hochschild G., Kostant B., Rosenberg A., Differential forms on
regular affine algebras, {\it Trans. Amer. Math. Soc.} {\bf 102} (1962), 383--408.

\bibitem{Ki92}
Kirwan F., Complex Algebraic Curves,
London Mathematical Society, Cambridge, 1992.

\bibitem{K97}
Kontsevich M., Deformation quantization of Poisson manifolds, I,
Preprint IHES (1997), q-alg/9709040.

\bibitem{Lo98}
Loday J.L., Cyclic homology, Springer-Verlag, Berlin, Heidelberg,
1998.

\bibitem{P05}
Pichereau A., Cohomologie de Poisson en dimension trois, {\it C.
R. Acad. Sci. Paris, Ser. I} {\bf 340} (2005).

\bibitem{P06}
Pichereau A., Poisson (co)homology and isolated singularities,
{\it Journal of Algebra} {\bf 299} (2006), Issue 2, 747--777.

\bibitem{RSP02}
Rannou E., Saux-Picart P., Cours de calcul formel, partie II,
\'editions Ellipses, 2002.

\bibitem{RV02}
Roger C., Vanhaecke P., Poisson cohomology of the affine plane,
{\it Journal of Algebra} {\bf 251} (2002), Issue 1, 448--460.

\bibitem{St93}
Sturmfels B., Algorithms in Invariant Theory, Texts and Monographs in Symbolic Computation, Springer Verlag, Wien, New-York, 1993.

\bibitem{VB94}
Van den Bergh M., Noncommutative homology of some
three-dimensional quantum spaces, in Proceedings of Conference on
Algebraic Geometry and Ring Theory
in honor of Michael Artin, Part III (Antwerp, 1992), volume 8, (1994), 213--230.\\


\end{thebibliography}
\end{document}